\def\wien{TUW--96/21}	\def\jhu{JHU-TIPAC-96020}   \def\tg{UTTG-15-96}

\documentclass[12pt]{article}    

\usepackage{amssymb,latexsym,cite,epic}	\long\def\del#1\enddel{}

        \input epsf

\let\3=\ss \catcode`\"=\active \let"=\"	\parindent=20pt	\parskip=6pt

\let\blackb=\mathbb		% \def\blackb#1{{\fam\black\relax#1}}	  
 \def\IZ{{\blackb Z}}   
\def\IR{{\blackb R}} \def\IC{{\blackb C}} \def\IP{{\blackb P}}  \let\BZ=\IZ

\let\a=\alpha \let\b=\beta   
  \let\th=\theta  
\let\l=\lambda \let\m=\mu \let\n=\nu   
\let\s=\sigma   \let\c=\chi
 
    \let\S=\Sigma \let\P=\Pi
  \let\G=\Gamma \let\D=\Delta
\def\0{\over }    \def\1{\vec }   \def\2{{1\over2}} \def\3{{\ss}}
\def\4{{1\over4}} \def\5{\overline }   \def\6{\partial } \def\7#1{{#1}\llap{/}}
\def\8#1{{\textstyle{#1}}}        \def\9#1{{\bf {#1}}}

\def\<{\langle } \def\>{\rangle }  
 
\def\({\left( } \def \){\right) }
 \let\then=\Rightarrow
\let\and=\wedge

\def\beq{\begin{equation}} \def\eeq{\end{equation}} \def\eeql#1{\label{#1}\eeq}
\def\bea{\begin{eqnarray}} \def\eea{\end{eqnarray}} 
\def\beqnn#1\eeq{\[#1\]}   \let\nn=\nonumber
\def\fnote#1#2{\begingroup\def\thefootnote{#1}\footnote{#2}
           \addtocounter{footnote}{-1}\endgroup}

\def\plb#1 #2 {Phys. Lett. {\bf B#1} #2 }
\def\phr#1 #2 {Phys. Rep. {\bf  #1} #2 } 
\def\npb#1 #2 {Nucl. Phys. {\bf B#1} #2 }
\def\aph#1 #2 {Ann. Phys. {\bf #1} #2 }  
\def\jmp#1 #2 {J. Math. Phys. {\bf #1} #2 }
\def\prd#1 #2 {Phys. Rev. {\bf D#1} #2 }
\def\prl#1 #2 {Phys. Rev. Lett. {\bf #1} #2 }
\def\rmp#1 #2 {Rev. Mod. Phys.  {\bf #1} #2 }
\def\zpc#1 #2 {Z. Phys. {\bf #1C} #2 }
\def\cmp#1 #2 {Commun. Math. Phys. {\bf #1} #2 }
\def\cqg#1 #2 {Class.Quant.Grav. {\bf #1} #2 }
\def\mpl#1 #2 {Mod. Phys. Lett. {\bf A#1} #2 }
\def\ijmp#1 #2 {Int. J. Mod. Phys. {\bf A#1} #2 }

 \def\ip{\hbox{\bf 1}} 
	\def\w{{\hbox{\footnotesize\bf k}}}
\def\ipo{\hbox{\bf 0}}		\def\w{{\bf k}}

\def\BP{\begin{picture}} \def\EP{\end{picture}}		%% --> PICTURE macros
\def\putlin#1,#2,#3,#4,#5){\put#1,#2){\line(#3,#4){#5}}} %\putlin(x,y,dx,dy,l)
\def\putvec#1,#2,#3,#4,#5){\put#1,#2){\vector(#3,#4){#5}}}
\def\putc#1)#2{\put#1){\makebox(0,0)[cc]{#2}}}
\def\BC{\begin{center}}  \def\EC{\end{center}}

\def\pmb#1{\setbox0=\hbox{${#1}$}   \kern-.025em\copy0\kern-\wd0
      \kern.05em\copy0\kern-\wd0     \kern-.025em\raise.0433em\box0 }

\def\putdot#1){\putc#1){\circle*2}}	\def\putnum#1)#2{\putc#1){\pmb{_#2}}}

\def\ifundefined#1{\expandafter\ifx\csname#1\endcsname\relax}

\newcounter{TRefNX} \let\OLDcite=\cite	\makeatletter%	     DRAFT MODE macros
\def\makeTRefs#1{\@for 	\NewTRef:=#1\do{\global\makeTRef{\NewTRef}}}
\def\makeTRef#1{\ifundefined{TRef#1}\stepcounter{TRefNX}%
\expandafter\xdef\csname TRef#1\endcsname{\theTRefNX}\fi}\makeatother
\def\NEWcite#1{\makeTRefs{#1}\OLDcite{#1}}  

\ifundefined{draftmode} {} \else	\let\cite=\NEWcite
\newcount\HOUR 	\HOUR=\the\time \divide\HOUR by 60	\multiply \HOUR by 60 
\newcount\MIN	\MIN=\the\time 	\advance\MIN by -\HOUR 	\divide\HOUR by 60
\ifnum	\MIN>9	\def\printTIME{{\it\the\HOUR\,:\,\the\MIN}}
	\else	\def\printTIME{{\it\the\HOUR\,:\,0\the\MIN}} \fi % \printTIME

\def\LLab#1{\BP(0,0)\unitlength=1mm\put(-12,.5){\makebox(0,0)[cr]{\small #1
        \rlap{$_{_{\makeatletter\csname TRef#1\endcsname\makeatother}}$}}}\EP}
\fi
\long\def\new#1\endnew{{\bf #1}}

\newfont{\XLbf}{cmbx10 scaled 2800}	\newfont{\XL}{cmr10 scaled 2600}
					\newfont{\xL}{cmr10 scaled 2100}

\begin{document}		
\baselineskip=16pt

{\hfill hep-th/9610154   \vskip1pt \hfill\tg \vskip1pt \hfill\wien 
\vskip 1pt\hfill\jhu}
\vskip 19mm

\centerline{\XL Searching for K3 Fibrations }		\vskip 9mm

\begin{center} \vskip 12mm 
	A.C. Avram${}^1$\fnote{a}{e-mail: alex@physics.utexas.edu},
       M. Kreuzer${}^2$\fnote{b}{e-mail: kreuzer@tph16.tuwien.ac.at},
	M. Mandelberg${}^3 $\fnote{c}{e-mail: isaac@bohr.pha.jhu.edu}
		              and
       H. Skarke${}^2$\fnote{d}{e-mail: skarke@tph16.tuwien.ac.at}
\vskip 5mm
  {\em ${}^1$Theory Group, Physics Department, 
              University of Texas at Austin\\
              Austin, TX 78712, USA\\[5pt]
       ${}^2$Institut f"ur Theoretische Physik, Technische Universit"at Wien\\
       Wiedner Hauptstra\3e 8--10, A-1040 Wien, AUSTRIA\\[5pt]
       ${}^3$The Johns Hopkins University, Baltimore, MD, 21218, USA}

\vfill                        {\bf ABSTRACT }
\end{center}

\noindent
We present two methods for studying fibrations of Calabi-Yau manifolds 
embedded in toric varieties described by single weight systems. 
We analyse 184,026 such spaces and identify among them 
124,701 which are K3 fibrations.
As some of the weights give rise to two or three distinct types of fibrations,
the total number we find is 167,406.
With our methods one can also study elliptic fibrations of 3-folds and K3 
surfaces.
We also calculate the Hodge numbers of the 3-folds
obtaining more than three times as many as were previously known.

\vfill \noindent %\wien\\[5pt] 
October 1996 \vspace*{9mm}
\thispagestyle{empty} \newpage
\pagestyle{plain} % \setcounter{page}{1}

\newpage

\section{Introduction}

Ever since Calabi-Yau manifolds entered into the lexicon of string theory, 
there have been various ways to classify them. 
Initially, it was (and, of course, still is) of great interest to 
calculate their Hodge numbers, thereby predicting the number of massless 
families of matter fields. 
This was of particular interest, since in the early days of the study of 
Calabi-Yau manifolds, there were few known examples, and so it seemed that 
it might be difficult to find one with Euler number $|\chi_E|=6$. 
By now there are dozens of topologically distinct Calabi-Yau manifolds 
with $|\chi_E|=6$, and in any case, these days we are happy to have 
manifolds with other values of $\chi_E$. 
They provide string theorists with a wide assortment of laboratories in 
which to test their ideas. 
\del [aa: i'd rather have this out. mm: why? hs: I agree with Alex.
What does this sentence tell us? ``Another'' as compared to what?]
Another {\it de facto} means of classifying Calabi-Yau manifolds was to 
distinguish them by how they were constructed. 
\enddel
In the beginning, there were  $\IP(4|5)$, $\IP(5|4~2)$, $\IP(5|3~3)$, 
$\IP(6|3~2~2)$, $\IP(7|2~2~2~2)$ and the $\IZ_3$ orbifold\cite{chsw}. 
Soon after followed complete intersections of polynomial hypersurfaces in 
direct products of projective spaces\cite{{cdls,yt}}, and then transverse 
polynomials in weighted projective spaces \cite{{nms,kl94}}. 
Most recently Calabi-Yau manifolds have been constructed as embeddings in 
toric varieties \cite{ba94}. 
Using this approach, it was shown in \cite{{crp,wtc}} that there are 
184,026 distinct sets of weights that define spaces in which Calabi-Yau 
manifolds can be embedded. 
These include both the weights that admit transverse polynomials, as well 
as those that don't, but nonetheless 
lead to reflexive polyhedra and thereby to varieties allowing
smooth Calabi-Yau hypersurfaces.
The purpose of this paper is to provide further means of 
analysing these manifolds. 

Among the dualities that have been recently studied is that between the 
$E_8\times E_8$ heterotic string compactified on $K3\times T^2$ and the 
Type IIA string compactified on a Calabi-Yau manifold. 
In \cite{kv} specific examples for this Calabi-Yau manifold were proposed. 
They rely on two pieces of evidence in support of the conjectured duality. 
The first is that the massless spectra on both sides match at generic points 
of their moduli spaces. 
The second involves a close examination of certain regions of the moduli 
space of the Calabi-Yau manifold. 
This analysis suggests a natural identification of one of the K\"{a}hler 
moduli with the heterotic dilaton. 
Specifically, the authors analyze the heterotic string with the $T^2$ chosen 
so that 
classically there is a single extra $SU(2)$ at the self-dual point, $\tau=i$. 
This suggests that when quantum effects are included, this singularity will 
split into two points where hypermulitplets become massless, much as in 
$N=2$ QFT \cite {sw}. 

As it turns out, there is a Calabi-Yau manifold that has the correct Hodge 
numbers. 
The only known Calabi-Yau manifold with $h_{11}=2$ and $h_{21}=128$ is 
{\cal M}=$\IP_4^{1,1,2,2,6}[12]$.
It is impressive that {\cal M} also has a moduli space that resembles the 
Seiberg-Witten solution. 
One of the keys to this correspondence comes from mirror symmetry.
By analysing the mirror map for {\cal M} $\leftrightarrow$ {\cal W}
\footnote{~   {\cal W} can be chosen as  $\IP_4^{12,13,18,25,28}[96]$.},
it can be shown that {\cal M} has a K\"{a}hler modulus that ``happens'' to 
have 
the modular transformation properties appropriate for a parameter describing 
the moduli space of an $N=2$ theory with $SU(2)$ gauge symmetry.
It was suggested in \cite{klm} that this is a generic feature of Calabi-Yau 
manifolds that are $K3$ fibrations. 
Indeed it is not difficult to see that $\IP_4^{1,1,2,2,6}[12]$ is a $K3$ 
fibration with standard fiber given by $\IP_3^{1,1,1,3}$. 
A short list of fibrations was provided in \cite{klm}. 
Of these, 31 are transverse hypersurfaces in $\IP_4^\w$.%
\footnote{~
	Another 25 are intersections of two hypersurfaces in $\IP_5^\w$.
} 
The authors proceed as follows. 
Consider a $K3$ given by a hypersurface in $\IP_3^\w$. 
If $\hbox{\bf k} = (1,k_2,k_3,k_4)$, then it is easy to see how to form a 
$K3$ fibration in $\IP_4^{1,1,2k_2,2k_3,2k_4}$. 
The $\IP_1$ base space is given by the ratio of the two coordinates with 
weight one, while the $K3$ is as above. 
Of the 95 $K3$'s that are given as a hypersurface in a $\IP_3^\w$, 
41 have a coordinate with weight one. 
Of these, 31 can generate a transverse polynomial in 
$\IP_4^\w$.%
\footnote{~
	If we allow for the $\IP_4^\w$'s described in \cite{{crp,wtc}}, 
	then this approach ``saturates the bound'', and all 41 
	of these $K3$'s produce fibrations.
} 
These results were extended in \cite{hly}. 
That article describes the general problem of determining whether a given 
$\IP_4^\w$ is a $K3$ fibration. 
The authors were able to identify 628 $K3$ fibrations among the 7,555 
$\IP_4^\w$ that admit transverse polynomials.

The question of why $K3$ fibrations are so important to heterotic-Type II 
duality has been largely laid to rest by Aspinwall and Louis in \cite{al}. 
Their result is that, assuming the Type IIA dual of the weakly coupled 
heterotic string is in the Calabi-Yau phase, then this Calabi-Yau manifold 
must be a $K3$ fibration. 
They achieve this result as a consequence of demanding that the 
pre-potentials for both sides match. 

Given this result, it has become useful to identify those Calabi-Yau 
manifolds that are $K3$ fibrations. 
We do this using the methods of toric geometry. 
In Section 2 we briefly review some elements of toric geometry. 
In Section 3 we describe in detail how a $K3$ fibration is manifested in a 
polyhedron. 
In \cite{cf} several such examples were analyzed. 
It was noted that in each case, the 
Newton polyhedron for the Calabi-Yau 
manifold exhibited the fibration in a particularly simple manner, to wit, 
there was a hyperplane through the origin whose intersection 
with the polyhedron gave a reflexive three dimensional subpolyhedron
\footnote{ ~
        It was first conjectured in \cite{klm} that the polyhedron of the 
        K3 should be a subpolyhedron of the polytope associated with the 
        Calabi-Yau family.
}. 
We will show that the general case is more complicated. 
Any projection of the 4 dimensional polyhedron onto a 3 dimensional 
sublattice that produces a reflexive polyhedron defines a $K3$ fibration. 
This is explained and demonstrated by examples. 
We also outline our approach for identifying fibrations. 
Section 4 provides the technical details. 
We explain how a certain kind of polyhedron known as a maximal 
Newton polyhedron has the property that if it has a face that is a 
reflexive 3 dimensional polyhedron, then there is always a 
unique projection onto that face.
We also explain our second strategy, which is somewhat complementary to the 
first, and 
also
relies on some simple geometric properties of how faces behave 
under projection. We summarize our results in Section 5. 
Of the 184,026 polyhedra that we analyze, we identify 124,701 that are 
$K3$ fibrations, 5,130 that are not, and 54,195 that do not yield to our 
methods. 
As an indication of the efficiency of our approaches, we note that
among the 7,555 transverse weights we find 5,370 models with one, two or
three fibrations 
(to be compared with the 628 fibrations of \cite{hly}).
Finally, we have calculated the Hodge numbers for all 184,026 polyhedra.
We find 14,121 different pairs of Hodge numbers, thereby
increasing the number of known pairs by a factor of more than three.

\section{Calabi--Yau hypersurfaces in toric varieties}

In toric geometry algebraic varieties are described with
the help of a dual pair of lattices $N$ and $M$, each isomorphic
to $\IZ^n$, and a fan $\S$ defined on $N_\IR$ (the real
extension of the $N$ lattice). 
Then there are various ways of constructing the variety $V_\S$ from
the toric data. 

In the old ``gluing'' approach, affine varieties 
$V_\s$ are associated with each cone $\s\in\S$, and $V_\S$
is obtained by gluing them together in a certain way (see, for example,
\cite{ful}).

In the holomorphic quotient approach \cite{cox} it is possible to assign
a single homogeneous coordinate system to $V_\S$ in a way
similar to the usual construction of $\IP^n$.
To this end one assigns a coordinate $z_k$, $k=1,\cdots,N$
(hopefully no confusion will arise from the two different
usages of the symbol $N$)
to each one dimensional cone in $\S$. 
If the primitive generators $v_1,\cdots,v_N$ of these one 
dimensional cones span $N_\IR$ (the real extension of $N$),
then there must be $N-n$ independent linear relations of
the type $\sum_k w^k_j v_k=0$. 
These linear relations are used to define equivalence relations
of the type 
\beq (z_1,\cdots,z_N)\sim (\l^{w^1_j}z_1,\cdots,\l^{w^N_j}z_N),~~~~~~
j=1,\cdots,N-n  \eeql{er} 
on the space $\IC^N-Z_\S$. 
The set $Z_\S$ is determined by the fan $\S$ in the following way:
It is the union of spaces 
$\{(z_1,\cdots,z_N):\; z_i=0\;\forall i\in I\}$,
where the index sets $I$ are those sets for which $\{v_i:\;i\in I\}$
does not belong to a cone in $\S$.
Thus $(\IC^*)^N\subset \IC^N \setminus Z_\S \subset \IC^N \setminus \{0\}$.
Then $V_\S=(\IC^N \setminus Z_\S)/(\IC^*)^{(N-n)}$, where the $N-n$ 
groups $\IC^*$
act by the equivalence relations given above.

A third approach is the symplectic quotient construction \cite{aud}, 
which is closely related to the holomorphic quotient construction. 
Here each of the $\IC^*$'s is decomposed as $\IR_+\times U(1)$.
One first chooses representatives of the $\IR_+^{N-n}$
equivalence classes by imposing $N-n$ equations 
$\sum_k w^k_j |z_k|^2=r_j$,
and then the resulting space is divided by the remaining $(U(1))^{N-n}$
invariance.

There are several reasons why we are mainly interested in the
holomorphic/symplectic quotient approach:
\begin{itemize}
\item The holomorphic quotient construction leads immediately to
the usual descriptions of projective spaces and is closely related
to weighted projective spaces.
\item Witten's gauged linear sigma model \cite{wit93} leads automatically 
to the symplectic quotient construction.
\item Describing Calabi--Yau spaces with Batyrev's method \cite{ba94}, 
the generators of one dimensional cones are just the integer points
of the dual polyhedron $\D^*$ in the $N$ lattice.
\item The description of K3 fibrations of Calabi--Yau spaces is
far easier with these constructions.
\end{itemize}

Let us briefly outline the construction of a Calabi--Yau
hypersurface in a space described by a reflexive polyhedron: 
We take $\D$ to be a reflexive polyhedron in $M_\IR$
(for example the Newton polyhedron of some weighted projective 
space), $\D^*\subset N_\IR$ its dual, and $\S$ a fan defined by
a maximal triangulation of $\D^*$.
This means that the integer generators $v_1,\cdots,v_N$ of the one 
dimensional cones are just the integer points (except the origin)
of $\D^*$. 
The polynomial that determines the CY--hypersurface 
takes the form
\beq \sum_{x\in \D\cap M}a_x\prod_{k=1}^N z_k^{\<v_k,x\>+1}.  \eeql{polyn}
It is easily checked that it is quasihomogeneous with respect to all 
$N-n$ relations of (\ref{er}) with degrees $d_j=\sum_{k=1}^N w_j^k$,
$j=1,\cdots N-n$.
Note how the reflexivity of the polyhedron ensures that the
exponents are nonnegative.

The Hodge numbers $h_{11}$ and $h_{21}$ for 3 dimensional
CY--hypersurfaces of this type are %% well known 
\cite{ba94} %% to be
\beq h_{11}=l(\D^*)-5-\sum_{{\rm codim }\th^* =1}l^*(\th^*)+
            \sum_{{\rm codim }\th^* =2}l^*(\th^*)l^*(\th)   \eeq
and
\beq h_{21}=l(\D)-5-\sum_{{\rm codim }\th =1}l^*(\th)+
            \sum_{{\rm codim }\th =2}l^*(\th^*)l^*(\th),   \eeq
where $\th$ and $\th^*$ are faces of $\D$ and $\D^*$, respectively,
and $l(\cdot)$ and $l^*(\cdot)$ denote the numbers of integer points
and integer interior points of polytopes.
These formulas are invariant under the simultaneous
exchange of $\D$ with $\D^*$ and $h_{11}$ with $h_{21}$
	so that Batyrev's construction is manifestly mirror symmetric
	(at least at the level of spectra).

\section{Toric K3 fibrations}

In the previous section we have seen how the polyhedron $\D^*$
determines the fan in the $N$ lattice which is used to describe
the ambient space, whereas the polyhedron $\D$ describes the 
polynomial whose vanishing locus is our CY hypersurface.
For the description of a fibration, we need the interplay 
between both sides. 
The following theorem relates properties of $\D$ and $\D^*$ which,
as we will soon see, are essential for the construction of
a fibration whose base space is $\IP^1$ and whose generic fiber is 
an $n-1$ dimensional CY space.
 
\noindent
{\bf Theorem:} Let $\D$ be an n dimensional reflexive polyhedron
in $M_\IR$. Then the following statements are
equivalent:\\
(1) There exists a projection operator $P:~~M\to M_{n-1}$,
where $M_{n-1}$ is an $n-1$ dimensional sublattice, such that
$P\D$ is a reflexive polyhedron in $M_{n-1}$.\\
(2) There is a lattice hyperplane in $N_\IR$ through the origin whose  
intersection with $\D^*$ is an $n-1$ dimensional reflexive polyhedron.\\
The respective $n-1$ dimensional polyhedra are dual to one another.
\\[4pt]
{\it Proof:} 
For showing that (1) implies (2), we choose a lattice basis 
$e_1,\cdots,e_n$ for $M$ such that $Pe_i=e_i$ for $i<n$, $Pe_n=0$
	(such a basis exists because $P$ projects onto a sublattice).
We denote coordinates with respect to this basis by $(x^i)$ 
and coordinates with respect to the dual basis $e^j$ of $N$ by $(y_j)$.
For (2) $\then$ (1) we start with a basis $e^j$ of $N$ such that
$\D^*\cap\{y_n=0\}$ is reflexive and define $e_i$ to be the dual basis.
In both cases we define 
\beq
\D_{n-1}=\{(x^1,\cdots,x^{n-1}):\exists~x^n \hbox{ with }
      (x^1,\cdots,x^{n-1},x^n)\in\D)\}
\eeq
and
\beq
(\D^*)_{n-1}=\{(y_1,\cdots,y_{n-1}): (y_1,\cdots,y_{n-1},0)\in\D^*\}.
\eeq
Now we note that we can prove both directions if we manage to
show that $\D_{n-1}$ and $(\D^*)_{n-1}$ are dual. 
This is the case, due to 
\bea
(\D_{n-1})^*&=&\{(y_1,\cdots,y_{n-1}): 
            \<(y_1,\cdots,y_{n-1}),(x^1,\cdots,x^{n-1})\>\ge -1\;
             \forall   \;(x^1,\cdots,x^{n-1})\in \D_{n-1}\}  \nn\\
    &=&\{(y_1,\cdots,y_{n-1}): 
            \<(y_1,\cdots,y_{n-1},0),(x^1,\cdots,x^n)\>\ge -1 \;
             \forall\; (x^1,\cdots,x^n)\in \D\}  \nn\\
    &=&(\D^*)_{n-1}.
\eea
\del
(1) $\then$ (2): Choose a lattice basis $e_1,\cdots,e_n$ for $M$
such that $Pe_i=e_i$ for $i<n$, $Pe_n=0$;
denote coordinates with respect to this basis by $(x^i)$.
Consider the prism consisting of points $(x^1,\cdots,x^{n-1},\l)$
($\l$ arbitrary)
such that there exists an $x^n$ with $(x^1,\cdots,x^{n-1},x^n)\in\D$.
Then $P$(prism)$=P\D=\D_{n-1}$. 
Reflexivity of $\D_{n-1}$ implies that the $n-2$ dimensional
bounding hyperplanes of $\D_{n-1}$ are described by equations
of the type $\<y,x\>=-1$, where the $y\in N_{n-1}$ (the lattice dual
to $M_{n-1}$) are the vertices of $\D_{n-1}^*$.
The bounding hyperplanes of the prism are described by 
$\<(y_1,\cdots,y_{n-1},0),x\>=-1$.
As duality inverts incidence relations, $\D\subset$ (prism)
implies that $\D^*$ contains all points of the form
$(y_1,\cdots,y_{n-1},0)$ with $(y_1,\cdots,y_{n-1})\in\D_{n-1}^*$.
If $\D^*\cap\{y_n=0\}$ contained points not in $\D_{n-1}^*$,
then $\D$ would have to lie in another prism (prism)' whose projection
would be a strict subset of the projection of the original prism, in
contradiction with $P\D=P$ (prism). Therefore   $\D^*\cap\{y_n=0\}$ 
must be isomorphic to $\D_{n-1}^*$.\\
(2) $\then$ (1): 
By going through the arguments of the first part of the proof in the
inverse direction.
\enddel
\hfill$\Box$\\[4pt]
This theorem can be readily generalized to the case where
$P:~M\to M_{n-k}$ is a projection on an $n-k$ dimensional
sublattice. In particular for $n=4$ and $k=2$ we can use it to study
elliptic fibrations of 3-folds.
The duality of projections and intersections is most easily
visualised with the help of a simple example:
Fig. 1 shows a dual pair of reflexive triangles. The first triangle 
has reflexive intersections along both coordinate axes, corresponding
to reflexive projections on the coordinate axes in the second triangle
(along the direction of the other axis).
The second triangle has only one reflexive intersection (along the
vertical axis), corresponding to the projection on the vertical
axis (along the horizontal axis) in the first triangle.

\begin{figure}		\ifundefined{epsfbox}

\BC	\unitlength=2.7pt % \def\putdot#1){\putc#1){\circle*3}}
\BP(40,60)(-20,-30)	\thicklines\put(43,-30){\makebox(0,0)[t]
{{\bf
	Fig. 1:} A dual pair of reflexive polygons
}}
	\putlin(-18,-18,1,0,36)\putlin(-18,-18,3,4,18)\putlin(18,-18,-3,4,18)
	\putdot(-18,-18)\putdot(-9,-18)\putdot(0,-18)\putdot(9,-18)
	\putdot(18,-18)
	\putdot(-18,-6)\putdot(-9,-6)\putdot(0,-6)\putdot(9,-6)\putdot(18,-6)
	\putdot(-18,6)\putdot(-9,6)\putdot(0,6)\putdot(9,6)\putdot(18,6)
			\thinlines
	\putlin(-27,-6,1,0,54) 
	\putlin(0,-27,0,1,42) 
\EP\hspace{42mm}
\BP(40,60)(-20,-30)	\thicklines
	\putlin(-18,-18,1,0,36)\putlin(-18,-18,3,4,18)\putlin(18,-18,-3,4,18)
	\putdot(-18,-18)	\putdot(0,-18)	\putdot(18,-18)
	\putdot(0,6) \putdot(0,-6)\putdot(-18,6) \putdot(-18,-6)
	\putdot(18,6) \putdot(18,-6)
			\thinlines
	\putlin(-27,-6,1,0,54) 
	\putlin(0,-27,0,1,42) 
\EP
\EC
\else
	\epsfxsize=12cm
	\hfil\epsfbox{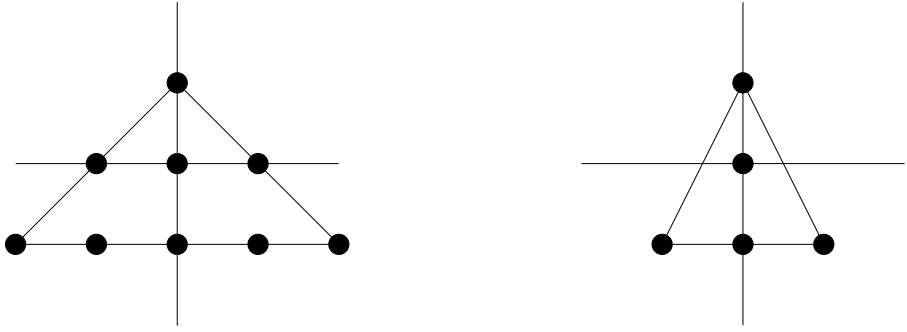}\hfill
	\caption{ A dual pair of reflexive polygons}
	\label{fig:pol}

\fi	\end{figure}

Let us now convince ourselves that reflexive polyhedra fulfilling
the criteria of the theorem correspond to spaces allowing
Calabi--Yau hypersurfaces that are fibrations whose generic
fibers are lower dimensional Calabi--Yau hypersurfaces.
Let us repeat our ingredients: In the $M$ lattice we have 
a distinguished direction with integer generator $e_n$
(as in the proof of the theorem) defining a projection $P$
that projects our polyhedron $\D$ to a reflexive
polyhedron $\D_{n-1}$.
In the $N$ lattice we have a distinguished hyperplane
\beq H=\{ y\in N: \<y,e_n\>=0\} \eeq
whose intersection with $\D^*$ is $\D_{n-1}^*$.

In the gluing approach we have the following picture: 
Depending on whether we consider $\D_{n-1}^*$ as a polytope in $N$ 
or in $N_{n-1}=N\cap H$, it defines $n$ or $n-1$ dimensional varieties 
related by
\beq V_{\D_{n-1}^*,N}\simeq V_{\D_{n-1}^*,N_{n-1}}\times \IC^*  \eeq
\del
In the gluing approach the toric variety defined by the subpolyhedron 
of $\D^*$ corresponding
to $\D_{n-1}^*$ can be identified with $\IC^*$ times the ambient
space of the lower dimensional Calabi--Yau hypersurface
\enddel
(see, for example, pages 5,6 of \cite{ful}).
Therefore $V_\D$ can be considered as a compactification of 
$V_{\D_{n-1}^*,N_{n-1}}\times \IC^*$, giving
a first hint in the desired direction.

A clearer picture emerges in the holomorphic quotient approach:
The integer generators $v_1,\cdots,v_N$ of the one dimensional cones 
are just the integer points of $\D^*$. 
We may split this set into $\{v_1,\cdots,v_{N'}\}$ (corresponding to 
points of $\D_{n-1}^*$) and the remaining set $\{v_{N'+1},\cdots,v_N\}$.
The latter set may again be decomposed, namely into the set of
points `above' and the set of points `below' $\D_{n-1}^*$.
Then we may choose our linear relations such that we have 
$N'-(n-1)$ relations involving only $v_1,\cdots,v_{N'}$
and $N-N'-1$ relations involving all of the $v_i$.
Similarly we may split our set $Z_{\Sigma_n}$ as
$Z_{\Sigma_n} \, = \, (Z_{\Sigma_{n-1}}\times \IC^{N-N'}) \cup Z_{\rm diff}$.
With a slight abuse of notation,
the embedding toric variety is then given by 
\beq \label{exsets}
\bigg\{ \bigg[ \Big[ \big( \IC^{N'} \setminus Z_{\Sigma_{n-1}} \big) 
\big/ 
{\IC^*}^{N'+1-n} \Big] \times
\IC^{N-N'} \bigg] \setminus Z_{\rm diff}  \bigg\} 
\bigg/ {\IC^*}^{N-N'-1}
\eeq
The direction of the projection $P$ determines a ray in the $M$ lattice
whose integer generator $e_n$ 
corresponds to a Laurent monomial 
\beq q(z_{N'+1},\cdots,z_N)=\prod_{j=1}^N z_j^{\<v_j,e_n\>} 
     =\prod_{j=N'+1}^N z_j^{\<v_j,e_n\>}. \eeq
We will now argue that the value of $q$ determines a point in our 
base space $\IP^1$.
A short glance at eq. (\ref{er}) shows us that $q$ is invariant
under the equivalence relations.
It takes the value $0$ if one of the $z_j$ corresponding
to points above $\D_{n-1}^*$ is $0$, and infinity if one of the 
$z_j$ corresponding to points below $\D_{n-1}^*$ is $0$.
If we choose our maximal triangulation of $\D^*$ in such a way
that it contains a maximal triangulation of $\D_{n-1}^*$,
the rules for constructing $Z_\S$ tell us that these two cases
cannot occur simultaneously.
Thus we have indeed a well--defined (and obviously surjective)
map from $V_\S$ to $\IP^1$.
The quasihomogeneous polynomial of eq. (\ref{polyn})
may be seen as a polynomial in $z_1,\cdots,z_{N'}$ 
(quasihomogeneous with respect to the first $N'-n+1$ relations)
with coefficients that are polynomials in $z_{N'+1},\cdots,z_N$.
After fixing a point in our base space $\IP^1$ we may then proceed
to determine the structure of the fiber.
In particular, if the point in the base space lies in $\IC^*\subset\IP^1$,
we may use the second set of relations to eliminate all
coordinates $z_{N'+1},\cdots,z_N$, thus obtaining the fiber
as the zero locus of a quasihomogeneous polynomial in the space
determined by $\D_{n-1}$. 
This shows us that the generic fiber in this construction is an
$n-2$ dimensional CY manifold.
A similar result is obtained at the level of embedding varieties.
Take a point where $q \neq 0$ is finite. Since $z_{N'+1},
\cdots,z_N$ are all different from zero there are no constraints
on the rest of the degrees of freedom coming from $Z_{\rm diff}$
which thus span the whole $V_{\Sigma_{n-1}}$.

How this works is best understood with the help of some examples.
The following two examples are based on 3 dimensional polyhedra
corresponding to elliptic fibrations of K3 manifolds,
because they can be visualised more easily. 
Nevertheless they exhibit all of the features that are essential also
in the 4 dimensional case.

\noindent
{\bf Example 1:}
Take as $\D^*$ the pyramid with peak at $v_1=(-1,-1,2)$ and
base the square whose vertices are $v_2=(0,0,-1),~~v_3=(1,1,-1),~~
v_4=(0,1,-1)$ and $v_5=(1,0,-1)$.
\ifundefined{epsfbox}\else
\begin{figure}
\epsfxsize=5cm
\hfil\epsfbox{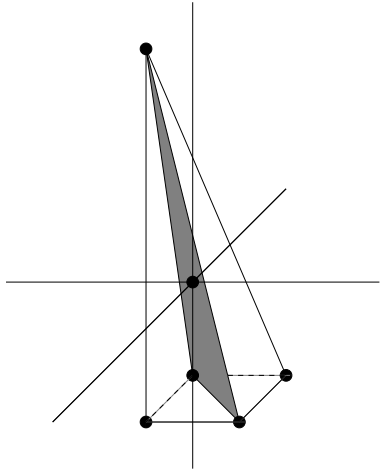}\hfill
\caption{The dual polytope $\D^*$ of Example 1}
\label{fig:dp1}
\end{figure}
\fi
\noindent
It is easily checked that there are no other integer points
(except the origin).
We have two linear relations: $v_1+v_2+v_3=0$ and $v_1+v_4+v_5=0$.
If we triangulate the base of the pyramid along the diagonal 
$\overline{v_2v_3}$, we get $Z_\S=\{z_1=z_2=z_3=0\}\cup\{z_4=z_5=0\}$.
With these data we can describe a 2--dimensional Calabi--Yau space
(a K3 surface) as a codimension 1 surface in $(\IC^5 \setminus 
Z_\S)/(\IC^*)^2$,
where the $(\IC^*)^2$ action is determined by 
\beq (z_1,z_2,z_3,z_4,z_5)\sim (\l\m z_1,\l z_2,\l z_3,\m z_4,\m z_5).
\eeq
The polynomial describing the surface is given by
\beq z_1^3+z_1^2(z_2p_1^{(1)}+z_3p_1^{(2)})+
     z_1(z_2^2p_2^{(1)}+z_2z_3p_2^{(2)}+z_3^2p_2^{(3)})+
     z_2^3p_3^{(1)}+z_2^2z_3p_3^{(2)}+z_2z_3^2p_3^{(3)}+z_3^3p_3^{(4)},
\eeq
where $p_1^{(.)},p_2^{(.)},p_3^{(.)}$ are linear, quadratic and
cubic polynomials in $z_4$ and $z_5$, respectively.
Clearly the whole polynomial is homogeneous of degree 3 both in $\l$
and in $\m$.
Our base space is $\IP^1$ with homogeneous coordinates $(z_4:z_5)$.
In the coordinate patch $z_4\ne 0$ ($z_5\ne 0$) we may use $\m$
to set $z_4$ ($z_5$) to 1.
Fixing a point in the base space amounts to fixing the values of
the $p_i^{(.)}$. The generic fiber is an elliptic curve
determined by a cubic equation in $\IP^2$ with homogeneous coordinates 
$(z_1:z_2:z_3)$.

\begin{figure}
\epsfxsize=5cm
\hfil\epsfbox{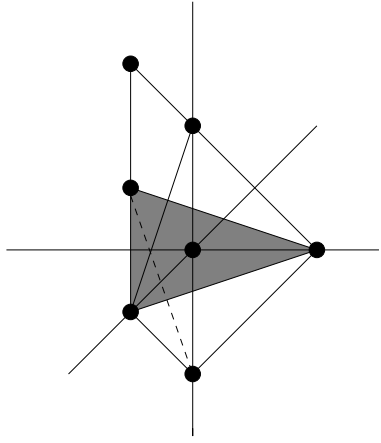}\hfill
\caption{The dual polytope $\D^*_3$ of Example 2}
\label{fig:dp2}
\end{figure}

\noindent
{\bf Example 2:}
Again we consider $N\simeq \BZ^3$. The polyhedron $\D^*_2$
corresponding to the fiber (the shaded area in Fig.\ref{fig:dp2}) 
is determined by the vertices
$v_1=(1,0,0),v_2=(0,1,0),v_3=(-1,-1,0)$, and the polyhedron
$\D^*_3$ has in addition two `upper' points $v_4=(0,0,1),v_5=(-1,-1,1)$
and one `lower'point $v_6=(0,0,-1)$.
\ifundefined{epsfbox}\else
\fi

\noindent

We have the linear relations $v_1+v_2+v_3=0$, $v_4+v_6=0$ and
$v_1+v_2+v_5+v_6=0$.
$\D^*_3$ has only triangular faces and
\beq
Z_\S=\{z_1=z_2=z_3=0\}\cup\{z_1=z_2=z_5=0\}\cup\{z_3=z_4=0\}\cup
\{z_4=z_6=0\}\cup\{z_5=z_6=0\}.
\eeq
$V_\S$ is the space $(\IC^6 \setminus Z_\S)/(\IC^*)^3$ with the $(\IC^*)^3$
action given by
\beq (z_1,z_2,z_3,z_4,z_5,z_6)\sim 
(\l\n z_1,\l\n z_2,\l z_3,\m z_4,\n z_5,\m\n z_6).
\eeq
We can describe a K3 surface in this space by an equation of the
type
\beq
z_1^3p_1^{(1)}+z_1^2z_2p_1^{(2)}+z_1z_2^2p_1^{(3)}+z_2^3p_1^{(4)}+
z_3(z_1^2p_2^{(1)}+z_1z_2p_2^{(2)}+z_2^2p_2^{(3)})+
z_3^2(z_1p_3^{(1)}+z_2p_3^{(2)})+z_3^3p_4=0
\eeq
where the $p_i^{(\cdot)}$ are polynomials in $z_4,z_5,z_6$ with bidegree
$(2,i)$ with respect to $(\m,\n)$.
The base space $\IP^1$ is $(z_4z_5:z_6)$.
In the patch where $z_4z_5\ne 0$ we may use $\m$ and $\n$
to set $z_4$ and $z_5$ to 1, thus taking $z_6$ as the variable 
parametrizing the $\IP^1$.
Again the generic fiber is an elliptic curve given by a cubic
equation in $\IP^2$ with homogeneous coordinates $(z_1:z_2:z_3)$.
At the point $z_4z_5=0$, however, the fiber becomes degenerate:
If $z_4=0$ then $z_3\ne 0$ and $z_6\ne 0$.
With suitable redefinitions of $\l,\m,\n$ and partial fixing
of the variables we obtain the $z_4=0$ part of the fiber as the
zero locus of a quadratic equation in a $\IP^2$ described by
\beq
(z_1,z_2,1,0,z_5,1)\sim (\n z_1,\n z_2,1,0,\n z_5,1),
\eeq
i.e. the $z_4=0$ part of the singular fiber is a double
cover of $\IP^1$.
At $z_5=0$ we have $(z_1,z_2)\ne (0,0)$. 
After suitable redefinitions of $\l,\m,\n$ we get an equation of the type
\beq
p_3(z_1,z_2)z_4+p_2(z_1,z_2)z_3=0
\eeq
in a space described by
\beq
(z_1,z_2,z_3,z_4,0,1)\sim (\a z_1,\a z_2,\a\b z_3,\b z_4,0,1).
\eeq
This space may be projected to a $\IP^1$ described by $(z_1:z_2)$.
Then the above equation determines $z_3$ and $z_4$ uniquely up
to $\b$ equivalence at each point $(z_1:z_2)$, so the $z_5=0$ part
of the fiber over $z_4z_5=0$ is just $\IP^1$. 
The two parts of the singular fiber intersect at the two points 
corresponding to the zero locus of a quadratic equation in $\IP^1$:
\beq
(z_1,z_2,1,0,0,1)\sim (\a z_1,\a z_2,1,0,0,1).
\eeq

Returning to our general discussion, we may ask whether our projection
$P\D$ in the $M$ lattice is isomorphic to the intersection of $\D$
with some lattice hyperplane $H$. If $H$ intersects the interior of
$\D$, then the only possibility for $\D\cap H$ to have an interior
point is that this point is the interior point of $\D$.
If $H$ does not intersect the interior of $\D$, then the requirement
that $\D\cap H$ is an $n-1$ dimensional polyhedron tells us
that $H$ must in fact be a bounding hyperplane, i.e. that $\D\cap H$ 
is a facet of $\D$. 
Summing up, there are three possibilities:
\begin{enumerate}
\item There is no hyperplane $H$ such that $P\D$ is isomorphic to $H\cap \D$;
\item $P\D$ is isomorphic to $H\cap \D$, where $H$ is a hyperplane through 
the origin;
\item $P\D$ is isomorphic to a facet of $\D$.
\end{enumerate}
Of course cases 2 and 3 are not mutually exclusive. 
An example is provided by the first triangle of Fig.1, with $P$
the projection along the horizontal axis.
There are also examples of polyhedra allowing different
projections to lower dimensional polyhedra, corresponding to different
cases. 
For the second triangle of Fig.1, the projection along the
horizontal axis corresponds to case 2 whereas the projection along the
vertical axis corresponds to case 3.
Case 2 is particularly pretty because there not only
the CY--manifold we are considering but also its mirror have
the structure of a fibration.
It is realised by the projections along the horizontal axes in Fig. 1.
Case 1 is realised, for example, by the triangle with vertices
(1,0), (0,1) and (-1,-1) (the dual of the Newton polyhedron of $\IP^2$).
Here the projections along both coordinate directions and along the 
(1,1)--direction are reflexive, but no intersection is.

Let us now consider possible strategies for looking for
torically realised K3 fibrations. 

Of course a complete analysis is possible by considering
all 3 dimensional hyperplanes in the $N$ lattice that are
linearly spanned by points of $\D^*$, one of which is the interior point. 
To this end, we would have to analyse ${l-1}\choose 3$ hyperplanes
for each polyhedron, where $l$ is the number of points in $\D$.
We did not attempt to do that.

A more modest approach is to search for projections corresponding
to case 3.
%Case 3 is the one where it is easiest to find examples. 
Hosono, Lian and Yau \cite{hly} have analysed the 7,555 CY--hypersurfaces
corresponding to transverse polynomials in weighted projective spaces
with respect to a specialisation of this case, where the reflexivity
of the facet can be read off from the weights.
In the present work we extend their results in two directions:
On the one hand we consider not only the 7,555 old models,
but the 184,026 models of \cite{wtc}, and on the other hand we analyse
the geometry of the faces instead of properties of weights,
thereby finding {\it all} fibrations corresponding to facets.

In fact, there is even more that we can learn by considering
facets of $\D$: 
Unless the projection $P$ is parallel to some facet $\th$,
any interior point of $\th$ will be mapped to an interior point 
of $P\D$. 
In particular, this implies the following:
If $\th$ has more than one interior point, $P\D$ can only be reflexive
if $P$ is parallel to $\th$, and if $\th$ has exactly one
interior point and $P\D$ is reflexive, then either $P$ is parallel 
to $\th$ or the interior point of $\th$ is mapped to the interior
point of $P\D$.
If $\D$ has enough facets with interior points, these considerations
are sufficient to determine all reflexive projections.

In the following section we will consider maximal Newton polyhedra
\cite{crp} corresponding to single weight systems (classified in
\cite{wtc}) and show how special properties of these polyhedra make the
application of these strategies particularly simple.

\section{Maximal Newton polyhedra}

In ref. \cite{crp}, an algorithm for the classification  of 
reflexive polyhedra was presented. The key to this algorithm is the 
fact that any reflexive polyhedron is a subpolyhedron of a polytope
defined by a weight system or a combination of weight systems, in the
following way:
A weight system is just a collection of positive rational numbers
$q_i=w_i/d$ with $\sum q_i=1$ ($\sum w_i=d$).
In the present work we consider only the case of a single
weight system with 5 weights. 
$(w_1,\cdots,w_5)$ defines a sublattice $\G^4$ of the lattice 
$\G^5\simeq\BZ^5$ by
\beq \G^4=\{(x^i)\in\G^5:\sum w_ix^i=d\}.    \eeql{g4}
If we also consider the real extensions $\G^5_\IR$ and $\G^4_\IR$,
we may define the simplex $Q$ (whose vertices are generically rational) 
as the
intersection of $\G^4_\IR$ with the positive hyperoctant in $\G^5_\IR$:
\beq Q=\{(x^i)\in\G^4_\IR: x^i\ge 0~~~ \forall i\}, \eeql{Q} 
and the maximal Newton polyhedron (MNP) $\D_{\rm max}$ as the set of 
integer points in $Q$: $\D_{\rm max}=Q\cap \G^4$. 
It is easy to see that the point $\ip=(1,\cdots,1)$ is the only
integer point in the interior of $Q$, and that its integer distance 
to any of the hyperplanes $x^i=0$ bounding $Q$ is 1.
If $\ip$ is also in the interior of $\D_{\rm max}$ (which we assume 
henceforth), we say that the weight system $(w_1,\cdots,w_5)$ has the 
`interior point property'. 
These weight systems, which play a crucial role
for the algorithm of \cite{crp}, were classified in \cite{wtc}.
In particular, in the case of 4 dimensions considered here, 
$\D_{\rm max}$ is reflexive by another result of \cite{wtc}.
The $M$ lattice is identified with $\G_4$, with $\ipo_M$ corresponding
to $\ip$.

It is tempting to identify spaces defined by an MNP 
with the weighted projective spaces defined by the weight system.
A weighted projective space, however, {\it is} a toric variety,
determined by a fan with n+1 one-dimensional cones. 
In our language, these are the cones over the vertices
of the integer (but not reflexive) simplex $Q^*$.
One can define a weighted projective space with any set of weights,
but the resulting variety is usually very ill behaved, whereas
the singularities of spaces defined by maximal triangulations of
reflexive polyhedra have codimension $\ge 4$ \cite{ba94}. 
What we consider are the varieties defined by the corresponding MNP's,
which are blowups of weighted projective spaces,
because $\D_{\rm max}\subset Q$ implies $Q^*\subset \D_{\rm max}^*$.
A special case occurs when the {\it vertices} of $Q^*$ are not just points,
but also {\it vertices} of $\D^*$. 
Then, by duality, the hyperplanes dual to these points 
(which are just the coordinate hyperplanes $x^i=0$ which carry the 
facets of the simplex $Q$) must also carry facets of $\D$.
Hence vertices of $Q^*$ are vertices of $\D^*$ if and only if 
the hyperplanes $x^i=0$ are affinely spanned by points of $\D$.
In this case we say that a weight system has the `span property'.
The weight systems with both the `interior point' and the `span property'
are the ones that are relevant for the classification scheme of \cite{crp}.

As we have seen in the previous section, interior points of
facets are the key ingredients for our strategies for the search
for reflexive projections of reflexive polyhedra.
The following lemmata will provide useful tools for identifying
interior points of facets of MNP's.
They all refer to MNP's defined by a single weight system.

\noindent
\del
{\bf Lemma 1:} 
Facets of $\D_{\rm max}$ that do not correspond to hyperplanes of the 
type $x^k=0$ have no interior points.
\\[4pt]
{\it Proof:} 
Such a point would have to be in the interior of $Q$, but the integer
interior point of $Q$ is unique and not on a facet.
\hfill$\Box$\\[4pt]
\enddel
{\bf Lemma 1:} 
An integer interior point of a facet $x^k=0$ of $Q$ defines a projection
$P:\;\G^4_\IR\to \G^4_\IR\cap\{x^k=0\}$ such that 
$P\D_{\rm max}=\D_{\rm max}\cap\{x^k=0\}$.
\\[4pt]
{\it Proof:} 
An integer interior point of a facet $x^k=0$ of $Q$ must have coordinates 
$x^k=0$ and $x^i=1+y^i,\;y^i\ge 0$
for $i\ne k$ (if one of the $x^i$ with $i\ne k$ were zero, our point would 
be at the boundary of the facet $x^k=0$). 
By comparison with the interior point
$(1,\cdots,1)$ of $\D_{\rm max}$ we see that $\sum_{i\ne k}w_iy^i=w_k$.
Then, for $(x^1,\cdots,x^5)\in \G^4_\IR$, $P$ is defined by $x^k\to 0$,
$x^i\to x^i+y^ix^k$ for $i\ne k$.
Obviously $\D_{\rm max}\cap\{x^k=0\}\subseteq P\D_{\rm max}$, and the 
maximality assumption on $\D_{\rm max}$ also ensures 
$P\D_{\rm max}\subseteq\D_{\rm max}\cap\{x^k=0\}$.
\hfill$\Box$\\[4pt]
{\bf Lemma 2:} 
An integer interior point of $Q\cap\{x^k=0\}$ is also an interior point of
$\D_{\rm max}\cap\{x^k=0\}$.
\\[4pt]
{\it Proof:} 
An integer interior point of $Q\cap\{x^k=0\}$ is the image of $(1,\cdots,1)$
under the projection $P$ of Lemma 1. 
If it were at the boundary of $\D_{\rm max}\cap\{x^k=0\}=P\D_{\rm max}$, 
then $(1,\cdots,1)$ would be at the boundary of $\D_{\rm max}$.
\hfill$\Box$\\[4pt]
{\bf Lemma 3:} 
(1) Facets of $\D_{\rm max}$ that do not correspond to hyperplanes of the 
type $x^k=0$ have no interior points.\\
(2) Interior points of a facet of $\D_{\rm max}$ defined by $x^k=0$
%has exactly one interior point if and only if it is 
%defined by $x^k=0$, where $w_k$ has a unique partition by $\{w_i:i\ne k\}$.
are in one-to-one correspondence with partitions of $w_k$ by $\{w_i:i\ne k\}$.
\\[4pt]
{\it Proof:} 
(1) Such a point would have to be in the interior of $Q$, but the integer
interior point of $Q$ is unique and not on a facet.\\
%Lemma 1 shows that we only need to consider facets of the type $x^k=0$.
(2) Partitions of $w_k$ by $\{w_i:i\ne k\}$ define projections as in the
proof of Lemma 2. Projections of $(1,\cdots,1)$ are integer interior 
points of $Q\cap\{x^k=0\}$ and (by Lemma 1) of $\D_{\rm max}\cap\{x^k=0\}$.
There is a one-to-one correspondence between partitions of $w_k$ and
interior points of $\D_{\rm max}\cap\{x^k=0\}$ defined by the corresponding 
projections.
\hfill$\Box$	%\\[4pt]

Lemma 3 is the key to our strategies for identifying reflexive projections:

Searching for reflexive facets, we simply look for weights that have 
unique partitions by the other weights. 
By Lemma 1 we can be sure that such a facet corresponds to a projection.
For 3 dimensional polyhedra (weight systems with 4 weights)
Lemma 3 would even guarantee reflexivity of the facet, because
any polygon with a single interior point is reflexive.
In the present case we still have to check for reflexivity
of the facet, and indeed it turns out that there are many cases
where a facet has a single interior point without being
reflexive.
If, for example, $w_1$ has the unique partition $w_1=\sum_{i>1}w_iy^i$,
then the vector $e_n$ determining our projection $P$ is given
by $(-1,y_2,\cdots,y_5)$.

On the other hand, if we have three facets (say, $x^1=0$, $x^2=0$
and $x^3=0$) with interior points, then any projection that does
not correspond to one of these facets must be parallel to each
of them.
Thus it must be parallel to the edge $x^1=x^2=x^3=0$ of $Q$,
connecting the vertices $(0,0,0,d/w_4,0)$ and $(0,0,0,0,d/w_5)$ 
of $Q$.
Denoting by $g$ the greatest common divisor of $w_4$ and $w_5$,
$e_n=(0,0,0,w_5/g,-w_4/g)$.
With four or five facets with interior points (this happens, in
particular, if we have one or two weights equal to 1), 
projections not corresponding to facets are excluded.

\section{Results}

%%%%%	local macros used for the tables:

	\def\nPro{\#$\P$}	\def\nFa{\#$F$}
	\def\VR#1#2{\vrule height #1mm depth #2mm width 0pt}
	\def\TVR#1#2{@{~~\VR{#1}{#2}}}		
	\def\NLHL{\\\hline}	
	\def\TN{T\,\bf{--}} \def\RS{{\bf{--}}\,S} \def\RN{\bf{--}\,\bf{--}}

In order to find K3 fibrations we pursued the two different strategies 
described in the previous sections. In the search for K3 facets we first
searched for weights that have a unique partition in the remaining weights,
which is a necessary condition for a unique interior point and hence for
reflexivity. The results are listed in Table 1 for the complete
list of reflexive weights and for the much smaller list of spanning weights.

\begin{center}					\def\UP{UP weight}
\smallskip
\begin{tabular}{||\TVR{4.5}2 c||c|c|c|c||c||} \hline\hline
weights		& no \UP  & 1 \UP  & 2 \UP s & 3 \UP s &
		total\\\hline\hline
reflexive	& 43,988 & 133,386 & 6,571 & 81 & 184,026\\\hline
spanning	& 9,939  & 26,400  & 2,361 & 27 & 38,727  \\\hline
\hline\end{tabular}
\\[4mm]
	{\bf Table 1:} Numbers of weights with unique partitions
\end{center}

If we keep only those weights whose corresponding facets actually are 
reflexive then the numbers reduce to the numbers of K3 fibrations that arise
as projections onto facets as given in Table 2.
It turns out that all of the 81 polytopes where 3 weights have
a unique partition have at most 1 reflexive facet. 
An example of this type is $(3,4,6,7,8;28)$, where 6, 7 and 
8 have unique partitions while only the
facet corresponding to the weight 7 is reflexive.

\del			% these are the spanning weights with  #UP=3 and #F=1 
7 3 4 6 8  28		7 3 4 8 9  31		10 3 4 8 9  34
8 3 5 9 10  35		11 3 5 9 10  38		9 4 5 12 15  45		
13 4 5 12 15  49	14 4 5 12 15  50	18 4 5 12 15  54	
11 5 6 18 20  60	21 5 6 18 25  75	21 4 9 20 27  81	
27 5 6 18 25  81	21 4 9 22 28  84	30 4 9 20 27  90	
21 4 9 28 31  93	51 4 15 42 56  168
\enddel

\vbox{
\begin{center}
\smallskip
\begin{tabular}{||\TVR{4.5}2 c||c|c|c||c||} \hline\hline
weights		& no K3 facet & 1 K3 facet & 2 K3 facets  &
		total\\\hline\hline
reflexive	& 76,460 & 104,036 & 3,530 & 184,026\\\hline
spanning	& 19,410 & 18,356  & 961  & 38,727 \\\hline
\hline\end{tabular}
\\[4mm]
	{\bf Table 2:} Numbers of K3 facets
\end{center}}

In both tables partitions of equal weights are counted only once because 
the corresponding projections onto facets are along the same 
direction $e_n=\pm(1,-1,0,\ldots)$ and thus define the same fibration
(regardless of the fact that there are two reflexive facets for each such pair 
of equal weights).
% in the case where the unique partition exists for a weight that occurs
% twice in the list this implies that there are 2 reflexive facets which,
% however, correspond to the same projection and hence to the same fibration!
This is always the case, for example, if 2 and only 2 weights are equal to 1 
(this type of example has been used extensively in the literature). 
A nice case with 2 pairs of equal weights is $(2,2,3,3,4;14)$, where 
2 and 3 have unique partitions, so there are 4 reflexive facets but
only 2 inequivalent K3 fibrations (see Table 6 below).

Our alternative search method, which is based on the constraints on directions 
of projection due to facets with interior points, is not applicable in 55\%
of all cases although about half of these even have projections onto K3
facets. In the remaining cases it allows, however, for a complete analysis,
providing many examples with up to 3 different K3 projections and a powerful
consistency check. In particular, for 5,130 weights we have shown that the
corresponding Newton polytopes do not admit any reflexive projection.
It is interesting to compare the results of our two approaches in some detail, 
and this is done by the 
compilations in the following tables, where we first give the 
overall statistics in Table 3, and then list the respective numbers for the
weights that do (T) or do not (--) admit transverse polynomials and for which
the Newton polytopes do (S) or do not (--) span all coordinate hyperplanes
in Table 4.
It turns out that the number of toric fibrations goes up to 3. Actually that
number can go up to at least 5, as one can see from the example of the 
mirror of the quintic threefold for which any projection along a line from
the origin to one of the 5 vertices is reflexive (since none of the facets has
an interior point, however, our algorithms do not find these fibrations).

\begin{figure}
\begin{center}				
%\begin{picture}(0,0)(0,0)\put(-50,-70){
%	{\bf Table 3:} Numbers of K3 projections (\nPro) and K3 facets (\nFa)}
%\end{picture}
%
\begin{tabular}{||c||c|c|c||c||} \hline\hline
	& \nFa=0& \nFa=1& \nFa=2& total	\\\hline\hline
\nPro=? & 54,195 & 45,792 & 1,508 & 101,495 \\\hline
\nPro=0 & 5,130 &  &  & 5,130 \\\hline
\nPro=1 & 17,135 & 20,080 &  & 37,215 \\\hline
\nPro=2 & 0 & 38,164 & 1,011 & 39,175 \\\hline
\nPro=3 & 0 & 0 & 1,011 & 1,011 \\\hline
\hline total & 76,460 & 104,036 & 3,530 & 184,026 \\\hline
\hline
\end{tabular}	
\\[4mm]{\bf Table 3:} Numbers of K3 projections (\nPro) and K3 facets (\nFa)
\end{center}\end{figure}

\begin{figure}
\begin{center}	\footnotesize	\small			
%\begin{picture}(0,0)(0,0)\put(5,-70)\end{picture}
%
\begin{tabular}{||c||c|c|c||c||} \hline\hline
TS	& \nFa=0& \nFa=1& \nFa=2& total	\\\hline\hline
\nPro=? & 703 & 299 & 12 & 1014 \\\hline
\nPro=0 & 507 &  &  & 507 \\\hline
\nPro=1 & 697 & 1,471 &  & 2,168 \\\hline
\nPro=2 & 0 & 383 & 88 & 471 \\\hline
\nPro=3 & 0 & 0 & 9 & 9 \\\hline
\hline total & 1,907 & 2,153 & 109 & 4,169 \\\hline
\hline
\end{tabular}
~
~
~
~
%\begin{picture}(0,0)(0,0)\put(0,-70){	}\end{picture}
\begin{tabular}{||c||c|c|c||c||} \hline\hline
\TN	& \nFa=0& \nFa=1& \nFa=2& total	\\\hline\hline
\nPro=? & 859 & 622 & 39 & 1,520 \\\hline
\nPro=0 & 116 &  &  & 116 \\\hline
\nPro=1 & 536 & 586 &  & 1,122 \\\hline
\nPro=2 & 0 & 577 & 27 & 604 \\\hline
\nPro=3 & 0 & 0 & 24 & 24 \\\hline
\hline total & 1,511 & 1,785 & 90 & 3,386 \\\hline
\hline
\end{tabular}
\\[5mm]	~~~~~{\bf Table 4a:} Transverse spanning weights
\hfill	{\bf Table 4b:} Transverse non-spanning weights ~ ~
\end{center}
%\vspace{5mm}
%			\end{figure}

%			\begin{figure}
\begin{center}\small
%\begin{picture}(0,0)(0,0)\put(5,-70){}\end{picture}
\begin{tabular}{||c||c|c|c||c||} \hline\hline
\RS	& \nFa=0& \nFa=1& \nFa=2& total	\\\hline\hline
\nPro=? & 10,222 & 3,804 & 297 & 14,323 \\\hline
\nPro=0 & 2,258 &  &  & 2,258 \\\hline
\nPro=1 & 5,023 & 6,692 &  & 11,715 \\\hline
\nPro=2 & 0 & 5,707 & 356 & 6,063 \\\hline
\nPro=3 & 0 & 0 & 199 & 199 \\\hline
\hline total & 17,503 & 16,203 & 852 & 34,558 \\\hline
\hline
\end{tabular}
~
~
~
%\begin{picture}(0,0)(0,0)\put(-5,-70){}\end{picture}
\begin{tabular}{||c||c|c|c||c||} \hline\hline
\RN	& \nFa=0& \nFa=1& \nFa=2& total	\\\hline\hline
\nPro=? & 42,411 & 41,067 & 1160 & 84,638 \\\hline
\nPro=0 & 2,249 &  &  & 2,249 \\\hline
\nPro=1 & 10,879 & 11,331 &  & 22,210 \\\hline
\nPro=2 & 0 & 31,497 & 540 & 32,037 \\\hline
\nPro=3 & 0 & 0 & 779 & 779 \\\hline
\hline total & 55,539 & 83,895 & 2,479 & 141,913 \\\hline
\hline
\end{tabular}
\\[5mm]	~~~{\bf Table 4c:} Non-transverse spanning weights	
\hfill		{\bf Table 4d:} Non-transverse non-spanning weights$\!$
\end{center}
%\vspace{9mm}
\end{figure}
%
%	15 17 18 19 26 95=d RN H:101 1 M:6 5 N:126 5 P:- F:0
%	51 60 64 65 80 320=d TN H:101 1 M:6 5 N:126 5 P:- F:0

In Table 5 we give the beginning and the end of our complete list of 
results which has 184,026 lines and is available via WWW 
\footnote{~
	The URL is http://tph.tuwien.ac.at/{\tiny$^\thicksim$}kreuzer/CY}
or by e-mail from the authors. 
Weights with unique partitions in terms of the other weights such that
the facets carried by the corresponding coordinate hyperplanes are reflexive 
are typed in bold face (if a weight occurs twice it is only marked once since
the pair defines only a single fibration; see above). We also list the 
Hodge numbers $h_{11}$ and $h_{21}$ and the numbers $P$ and $V$ of points and
vertices for the MNP and its dual, and also indicate the transversality and
spanning properties of the weights.

\begin{figure}
\begin{center}		
% \vspace*{-3mm}
\begin{tabular}{||\TVR{3.5}1 c|ccccc|c|cc|cc|cc|cc||} \hline\hline	\VR41
$d$	&$w_1$	& $w_2$	& $w_3$ & $w_4$ & $w_5$ &  TS	& $h_{11}$ & $h_{12}$ 
& $\#P$	& $\#V$	& $\#\5P$	& $\#\5V$	& $\P$	& F \\\hline\hline
5 &1 & 1 & 1 & 1 & 1 & TS& 1&101 & 126&5 & 6&5 & 0&0\NLHL
6 &1 & 1 & 1 & 1 & 2 & TS& 1&103 & 130&5 & 6&5 & 0&0\NLHL
7 &1 & 1 & 1 & 1 & 3 & TS& 2&122 & 159&8 & 7&6 & 0&0\NLHL
7 &1 & 1 & 1 & 2 & 2 & TS& 2&95 & 120&9 & 7&6 & 0&0\NLHL
8 &1 & 1 & 1 & 1 & 4 & TS& 1&149 & 201&5 & 7&5 & 0&0\NLHL
8 &1 & 1 & 1 & 2 & 3 & TS& 2&106 & 136&8 & 7&6 & 0&0\NLHL
8 &\bf1 & 1 & 2 & 2 & 2 & TS& 2&86 & 105&5 & 7&5 & 1&1\NLHL
9 &1 & 1 & 1 & 2 & 4 & TS& 3&123 & 162&9 & 8&6 & 0&0\NLHL
9 &1 & 1 & 1 & 3 & 3 & TS& 4&112 & 145&5 & 7&5 & 0&0\NLHL
9 &1 & 1 & 2 & 2 & 3 & TS& 2&86 & 109&9 & 7&6 & 0&0\NLHL
10 &1 & 1 & 1 & 2 & 5 & TS& 1&145 & 196&5 & 7&5 & 0&0\NLHL
10 &1 & 1 & 1 & 3 & 4 & \RS& 4&126 & 165&10 & 9&7 & 0&0\NLHL
10 &\bf1 & 1 & 2 & 2 & 4 & TS& 3&99 & 126&8 & 8&6 & 1&1\NLHL
10 &1 & 1 & 2 & 3 & 3 & TS& 3&87 & 111&9 & 8&6 & 0&0\NLHL
10 &1 & 2 & 2 & 2 & 3 & TS& 3&75 & 87&8 & 8&6 & 0&0\NLHL
11 &1 & 1 & 1 & 3 & 5 & TS& 4&144 & 192&10 & 9&7 & 0&0\NLHL
11 &\bf1 & 1 & 2 & 2 & 5 & TS& 4&109 & 144&10 & 9&7 & 1&1\NLHL
11 &1 & 1 & 2 & 3 & 4 & TS& 4&94 & 121&13 & 9&8 & 0&0\NLHL
11 &1 & 2 & 2 & 3 & 3 & TS& 4&64 & 81&13 & 9&7 & 0&0\NLHL
12 &1 & 1 & 1 & 3 & 6 & TS& 3&165 & 225&5 & 8&5 & 0&0\NLHL
12 &1 & 1 & 1 & 4 & 5 & \RS& 4&154 & 204&7 & 10&6 & 0&0\NLHL
12 &\bf1 & 1 & 2 & 2 & 6 & TS& 2&128 & 171&5 & 8&5 & 1&1\NLHL
12 &1 & 1 & 2 & 3 & 5 & TS& 3&105 & 137&7 & 8&6 & 0&0\NLHL
12 &\bf1 & 1 & 2 & 4 & 4 & TS& 5&101 & 130&5 & 8&5 & 1&1\NLHL
12 &1 & 1 & 3 & 3 & 4 & TS& 5&89 & 115&5 & 7&5 & 0&0\NLHL
12 &1 & 2 & 2 & 2 & 5 & TS& 4&94 & 108&8 & 9&6 & 0&0\NLHL
12 &1 & 2 & 2 & 3 & 4 & TS& 2&74 & 89&5 & 7&5 & 0&0\NLHL
12 &1 & \bf2 & 3 & 3 & 3 & TS& 3&69 & 81&5 & 8&5 & 1&1\NLHL
12 &2 & 2 & 2 & \bf3 & 3 & TS& 6&60 & 63&5 & 8&5 & 1&1\NLHL
	   $\cdots$ & $\cdots$ &$\cdots$ &$\cdots$ &$\cdots$ &$\cdots$ &
$\cdots$ & $\cdots$ & $\cdots$ &$\cdots$ &$\cdots$ &$\cdots$ &$\cdots$ &
						    $\cdots$ &$\cdots$ \NLHL
3192 &37 & 39 & \bf456 & 1064 & 1596 & \RN& 491&11 & 26&5 & 680&5 & 2&1\NLHL
3234 &36 & 41 & \bf462 & 1078 & 1617 &\TN& 462&12 & 27&6 & 639&6 & 2&1\NLHL
3234 &37 & 40 & \bf462 & 1078 & 1617 & \RN& 491&11 & 26&5 & 680&5 & 2&1\NLHL
3234 &38 & 39 & \bf462 & 1078 & 1617 & \RN& 491&11 & 26&5 & 680&5 & 2&1\NLHL
3276 &37 & 41 & \bf468 & 1092 & 1638 & \RN& 491&11 & 26&5 & 680&5 & 2&1\NLHL
3318 &37 & 42 & \bf474 & 1106 & 1659 & \RN& 491&11 & 26&5 & 680&5 & 2&1\NLHL
3318 &38 & 41 & \bf474 & 1106 & 1659 & \RN& 491&11 & 26&5 & 680&5 & 2&1\NLHL
3318 &39 & 40 & \bf474 & 1106 & 1659 & \RN& 491&11 & 26&5 & 680&5 & 2&1\NLHL
3360 &39 & 41 & \bf480 & 1120 & 1680 & \RN& 491&11 & 26&5 & 680&5 & 2&1\NLHL
3402 &40 & 41 & \bf486 & 1134 & 1701 & \RN& 491&11 & 26&5 & 680&5 & 2&1\NLHL
3486 &41 & 42 & \bf498 & 1162 & 1743 &\TN& 491&11 & 26&5 & 680&5 & 2&1\NLHL
\hline\end{tabular}
\\[4mm]
	{\bf Table 5:} Results from the complete list of reflexive weights
	(available via WWW)
\end{center}
\end{figure}

In Table 6 we give the same information for a number of examples that
illustrate the phenomena that can occur. The first line is an example
of a weight combination whose MNP is the mirror of the quintic, i.e. a
simplex with 6 points such that all vertices have distance 5 from the  
opposite facet. This polytope has 5 reflexive projections, but none is onto
a facet and it also cannot be analysed with our second approach since none
of the facets has an interior point. The second example features the maximal
number of vertices of $\D$, which is 18; $\D^*$ can have up to 21 vertices,
which occurs twice. The number of points goes up to 680 both for $\D$ 
(this number occurs for the Fermat weights in the next line of Table 6) 
and $\D^*$ (apparently the dual of the Fermat simplex, represented many
times by different weight systems, for example at the end of Table 5). 
The remaining lines
give examples (of minimal degree) for all combinations of transversality and
spanning properties, general projections and projections onto facets.

%%%%%%%  elliptic K3s
	It is straightforward to apply our methods also to the well known
	95 K3 weights, this time producing elliptic fibrations. Since 
	all 2D polytopes with one interior point are automatically reflexive,
	no reflexivity check is required 
	and the search for `elliptic' facets can be done by hand.
	The results are listed in Table~7, where the (up to 2) weights with 
	unique partition are again in bold face. With our alternative search
	algorithm we again find up to 3 fibrations per Newton polytope, 
	so that we find a total of 110 ellipitic fibrations.%
\footnote{~
	The algorithm of \cite{hly} yields 18, 
	with 1 double fibration.
}
	The statistics are given in Table~8. 

We also computed the Hodge numbers and compared and combined them with
the complete results that exist for weighted projective spaces \cite{nms,kl94}
and abelian orbifolds thereof \cite{ao} in Table~9.
A pronounced feature is the absence of mirror symmetry, which is already 
familiar from the list of transversal weights. 
	The well known plot Fig.4 becomes much denser,
	but otherwise does not change shape.
In our context
it does not make sense any longer to omit the mirror spectra since they are
produced by the dual polytopes (which, however, cannot be MNPs for a 
single weight system whenever the spectrum does not occur in our original
list).
Counting all spectra of MNPs and their duals we thus get an increase in the
known Calabi--Yau spectra by more than a factor of 3. Inclusion of the 
abelian orbifold spectra for the transversal weights,
which have been analysed completely in \cite{ao},
only adds 173 new spectra to those for reflexive
MNPs of single weight systems. 
If we mirror-symmetrize by hand, as is appropriate in the
toric framework (all abelian orbifolds correspond to MNPs on sublattices,
which are reflexive because of the results of \cite{wtc}), this number
even goes down to 95.
This can be interpreted as an indication that our list of spectra already
might be fairly complete.

\begin{figure}
\begin{center}		
% \vspace*{-3mm}
\begin{tabular}{||\TVR{3.5}1 c|ccccc|c|cc|cc|cc|cc||} \hline\hline	\VR41
$d$	&$w_1$	& $w_2$	& $w_3$ & $w_4$ & $w_5$ &  TS	& $h_{11}$ & $h_{12}$ 
& $\#P$	& $\#V$	& $\#\5P$	& $\#\5V$	& $\P$	& F \\\hline\hline
95 &15 & 17 & 18 & 19 & 26 & \RN& 101&1 & 6&5 & 126&5 & ?&0\\\hline
47 &3 & 4 & 5 & 14 & 21 & \RS& 26&39 & 54&\bf18 & 35&15 & ?&0\\\hline
69 &7 & 8 & 10 & 19 & 25 & \RS& 59&10 & 16&13 & 75&\bf21 & ?&0\\
97 &7 & 8 & 11 & \bf26 & 45 & \RS& 63&15 & 24&15 & 71&\bf21 & ?&1\\\hline
84 &\bf1 & 1 & 12 & 28 & 42 & TS& 11&491 & 680&5 & 26&5 & 1&1\\\hline
% 70 &\bf1 & 1 & 10 & 23 & 35 & TS& 14&416 & \bf576&6 & 26&6 & 1&1\\\hline
% 280 &7 & 19 & \bf40 & 87 & 127 & \RN& 491&11 & 26&5 & \bf680&5 & 2&1\\\hline
24 &3 & 4 & 5 & 6 & 6 & TS& 10&34 & 36&8 & 12&7 & ?&0\\
26 &3 & 4 & 5 & 7 & 7 & \RS& 22&22 & 31&13 & 21&10 & ?&0\\
33 &3 & 6 & 6 & 7 & 11 & \RN& 19&37 & 34&7 & 22&6 & ?&0\\
36 &3 & 6 & 6 & 10 & 11 & \TN& 19&49 & 38&7 & 22&6 & ?&0\\\hline
26 &3 & 4 & 5 & \bf6 & 8 & \RS& 14&24 & 32&14 & 19&10 & ?&1\\
36 &5 & \bf7 & 7 & 8 & 9 & \RN& 30&12 & 19&10 & 28&9 & ?&1\\
39 &3 & \bf6 & 9 & 10 & 11 & TS& 17&41 & 33&12 & 22&13 & ?&1\\
52 &4 & 6 & \bf8 & 11 & 23 & \TN& 29&33 & 34&9 & 36&8 & ?&1\\\hline
34 &3 & \bf6 & 7 & 8 & \bf10 & \RS& 18&20 & 27&13 & 23&12 & ?&2\\
44 &4 & \bf8 & 9 & 10 & \bf13 & \RN& 29&17 & 22&9 & 31&9 & ?&2\\
55 &3 & 10 & \bf13 & 14 & \bf15 & TS& 28&16 & 23&12 & 35&14 & ?&2\\
63 &7 & 9 & \bf14 & 15 & \bf18 & \TN& 44&8 & 15&6 & 37&6 & ?&2\\\hline
5 &1 & 1 & 1 & 1 & 1 & TS& 1&101 & 126&5 & 6&5 & 0&0\\
10 &1 & 1 & 1 & 3 & 4 & \RS& 4&126 & 165&10 & 9&7 & 0&0\\
25 &1 & 5 & 5 & 6 & 8 & \TN& 17&49 & 65&7 & 15&7 & 0&0\\
26 &1 & 5 & 5 & 7 & 8 & \RN& 19&49 & 65&9 & 19&7 & 0&0\\\hline
20 &2 & 3 & 4 & 4 & 7 & \RS& 13&45 & 51&10 & 14&8 & 1&0\\
20 &2 & 3 & 5 & 5 & 5 & TS& 6&48 & 50&8 & 11&6 & 1&0\\
30 &2 & 5 & 6 & 6 & 11 & \RN& 27&39 & 45&7 & 25&6 & 1&0\\
36 &2 & 5 & 6 & 6 & 17 & \TN& 24&54 & 60&7 & 25&6 & 1&0\\\hline
8 &\bf1 & 1 & 2 & 2 & 2 & TS& 2&86 & 105&5 & 7&5 & 1&1\\
13 &\bf1 & 1 & 2 & 4 & 5 & \RS& 6&108 & 141&12 & 11&8 & 1&1\\
35 &2 & 7 & \bf8 & 9 & 9 & \RN& 35&23 & 33&11 & 30&8 & 1&1\\
40 &4 & 5 & \bf9 & 10 & 12 & \TN& 22&18 & 25&7 & 20&7 & 1&1\\\hline
19 &2 & 3 & \bf4 & 5 & 5 & \RS& 11&33 & 43&14 & 14&9 & 2&1\\
27 &2 & 3 & \bf4 & 9 & 9 & TS& 14&44 & 56&9 & 13&7 & 2&1\\
36 &\bf4 & 4 & 6 & 9 & 13 & \RN& 31&31 & 33&6 & 29&6 & 2&1\\
40 &\bf4 & 4 & 6 & 9 & 17 & \TN& 26&38 & 39&7 & 29&6 & 2&1\\\hline
14 &\bf2 & 2 & \bf3 & 3 & 4 & TS& 5&51 & 57&10 & 10&7 & 2&2\\
19 &2 & \bf3 & 3 & \bf4 & 7 & \RS& 11&39 & 51&14 & 16&9 & 2&2\\
30 &3 & \bf5 & 5 & \bf6 & 11 & \RN& 33&21 & 33&6 & 25&6 & 2&2\\
35 &3 & \bf5 & 5 & \bf6 & 16 & \TN& 26&28 & 42&7 & 25&6 & 2&2\\\hline
28 &4 & \bf5 & 5 & 6 & \bf8 & \RS& 18&20 & 27&10 & 18&8 & 3&2\\
36 &4 & 6 & \bf8 & \bf9 & 9 & TS& 23&23 & 26&6 & 16&6 & 3&2\\
40 &4 & \bf7 & 7 & 10 & \bf12 & \RN& 28&16 & 23&7 & 25&6 & 3&2\\
42 &6 & \bf7 & 7 & 10 & \bf12 & \TN& 35&11 & 19&6 & 23&6 & 3&2\\
\hline\end{tabular}
\\[4mm]
	{\bf Table 6:} 	Examples from the long list of weights with different
			features
\end{center}
\end{figure}

\begin{figure}	\vspace{-9mm}
\begin{center}	{	\footnotesize	\def\NLHL{\\}	\def\N{--}
% \vspace*{-3mm}
\def\tabtop
{\begin{tabular}{||\TVR{3.5}1 
		c|c@{~}c@{~}c@{~}c|c|c@{~}c@{~}|c@{~}c@{~}|c@{~~~}c||} 
	\hline\hline	\VR41
$d$	&$w_1$	& $w_2$	& $w_3$ & $w_4$ &  S	
	& $\#P$	& $\#V$	& $\#\5P$ & $\#\5V$ & $\P$	
	& F \\\hline\hline
}							\hspace*{-6mm}\tabtop
4 & 1 &1 &1 &1 & S& 35 & 4 & 5 & 4 & 0 & 0\NLHL
5 & 1 &1 &1 &2 & S& 34 & 6 & 6 & 5 & 0 & 0\NLHL
6 & 1 &1 &1 &3 & S& 39 & 4 & 6 & 4 & 0 & 0\NLHL
6 & \bf1 &1 &2 &2 & S& 30 & 4 & 6 & 4 & 1 & 1\NLHL
7 & \bf1 &1 &2 &3 & S& 31 & 7 & 8 & 6 & 1 & 1\NLHL
8 & 1 &2 &2 &3 & S& 24 & 6 & 8 & 5 & 0 & 0\NLHL
8 & \bf1 &1 &2 &4 & S& 35 & 4 & 7 & 4 & 1 & 1\NLHL
9 & 1 &\bf2 &3 &3 & S& 23 & 6 & 8 & 5 & 1 & 1\NLHL
9 & \bf1 &1 &3 &4 & S& 33 & 5 & 9 & 5 & 1 & 1\NLHL
10 & 1 &2 &2 &5 & S& 28 & 4 & 8 & 4 & 0 & 0\NLHL
10 & 1 &\bf2 &3 &4 & S& 23 & 7 & 11 & 6 & 1 & 1\NLHL
10 & \bf1 &1 &3 &5 & S& 36 & 5 & 9 & 5 & 1 & 1\NLHL
11 & 1 &\bf2 &3 &5 & S& 24 & 8 & 13 & 7 & 1 & 1\NLHL
12 & 1 &\bf2 &3 &6 & S& 27 & 4 & 9 & 4 & 1 & 1\NLHL
12 & 1 &\bf2 &4 &5 & S& 24 & 5 & 12 & 5 & 1 & 1\NLHL
12 & 1 &\bf3 &4 &4 & S& 21 & 4 & 9 & 4 & 1 & 1\NLHL
12 & 2 &\bf3 &3 &\bf4 & S& 15 & 4 & 9 & 4 & 2 & 2\NLHL
12 & \bf1 &1 &4 &6 & S& 39 & 4 & 9 & 4 & 1 & 1\NLHL
12 & \bf2 &2 &3 &5 & S& 17 & 5 & 11 & 5 & 1 & 1\NLHL
13 & 1 &\bf3 &4 &5 & S& 20 & 7 & 15 & 7 & 1 & 1\NLHL
14 & 1 &\bf2 &4 &7 & S& 27 & 5 & 12 & 5 & 1 & 1\NLHL
14 & 2 &3 &\bf4 &\bf5 & S& 13 & 7 & 16 & 7 & 3 & 2\NLHL
14 & \bf2 &2 &3 &7 & S& 19 & 5 & 11 & 5 & 1 & 1\NLHL
15 & 1 &\bf2 &5 &7 & S& 26 & 6 & 17 & 6 & 1 & 1\NLHL
15 & 1 &\bf3 &4 &7 & S& 22 & 6 & 17 & 6 & 1 & 1\NLHL
15 & 1 &\bf3 &5 &6 & S& 21 & 5 & 15 & 5 & 1 & 1\NLHL
15 & 2 &3 &5 &5 & S& 14 & 6 & 11 & 5 & 1 & 0\NLHL
15 & \bf3 &3 &4 &5 & S& 12 & 5 & 12 & 5 & 1 & 1\NLHL
16 & 1 &\bf2 &5 &8 & S& 28 & 5 & 14 & 5 & 1 & 1\NLHL
16 & 1 &\bf3 &4 &8 & S& 24 & 5 & 12 & 5 & 1 & 1\NLHL
16 & 1 &\bf4 &5 &6 & S& 19 & 6 & 17 & 6 & 1 & 1\NLHL
16 & 2 &3 &\bf4 &7 & S& 14 & 6 & 18 & 6 & 2 & 1\NLHL
17 & 2 &3 &\bf5 &7 & S& 13 & 8 & 20 & 8 & 2 & 1\NLHL
18 & 1 &\bf2 &6 &9 & S& 30 & 4 & 12 & 4 & 1 & 1\NLHL
18 & 1 &\bf3 &5 &9 & S& 24 & 5 & 15 & 5 & 1 & 1\NLHL
18 & 1 &\bf4 &6 &7 & S& 19 & 6 & 20 & 6 & 1 & 1\NLHL
18 & 2 &3 &\bf4 &9 & S& 16 & 5 & 14 & 5 & 2 & 1\NLHL
18 & 2 &3 &\bf5 &8 & S& 14 & 6 & 20 & 6 & 2 & 1\NLHL
18 & 3 &4 &5 &\bf6 & S& 10 & 6 & 17 & 6 & ? & 1\NLHL
19 & 3 &4 &5 &\bf7 & S& 9 & 7 & 24 & 8 & ? & 1\NLHL
20 & 1 &\bf4 &5 &10 & S& 23 & 4 & 13 & 4 & 1 & 1\NLHL
20 & 2 &3 &\bf5 &10 & S& 16 & 5 & 14 & 5 & 2 & 1\NLHL
20 & 2 &5 &\bf6 &\bf7 &\N& 11 & 5 & 23 & 5 & 3 & 2\NLHL
20 & 2 &\bf4 &5 &9 &\N& 13 & 4 & 23 & 4 & 1 & 1\NLHL
20 & 3 &4 &5 &8 & S& 10 & 6 & 22 & 6 & ? & 0\NLHL
21 & 1 &\bf3 &7 &10 &\N& 24 & 4 & 24 & 4 & 1 & 1\NLHL
21 & 1 &\bf5 &7 &8 &\N& 18 & 5 & 24 & 5 & 1 & 1\NLHL
21 & 2 &3 &\bf7 &9 & S& 14 & 6 & 23 & 6 & 2 & 1\NLHL
\hline\end{tabular}	\hspace{9mm}
%
% \begin{tabular}{||\TVR{3.5}1 c|cccc|c|cc||} \hline\hline	\VR41
% $d$	&$w_1$	& $w_2$	& $w_3$ & $w_4$ &  S	& $\P$	& F \\ \hline\hline
	\tabtop
21 & 3 &5 &\bf6 &7 &\N& 9 & 5 & 21 & 5 & ? & 1\NLHL
22 & 1 &\bf3 &7 &11 &\N& 25 & 5 & 20 & 5 & 1 & 1\NLHL
22 & 1 &\bf4 &6 &11 & S& 22 & 6 & 20 & 6 & 1 & 1\NLHL
22 & 2 &\bf4 &5 &11 &\N& 14 & 5 & 19 & 5 & 1 & 1\NLHL
24 & 1 &\bf3 &8 &12 & S& 27 & 4 & 15 & 4 & 1 & 1\NLHL
24 & 1 &\bf6 &8 &9 & S& 18 & 5 & 24 & 5 & 1 & 1\NLHL
24 & 2 &3 &8 &11 &\N& 15 & 4 & 27 & 4 & 1 & 0\NLHL
24 & 2 &3 &\bf7 &12 & S& 16 & 5 & 20 & 5 & 2 & 1\NLHL
24 & 3 &4 &5 &12 & S& 12 & 5 & 18 & 5 & ? & 0\NLHL
24 & 3 &4 &\bf7 &10 & S& 10 & 5 & 26 & 6 & 2 & 1\NLHL
24 & 3 &\bf6 &7 &8 &\N& 9 & 4 & 21 & 4 & ? & 1\NLHL
24 & 4 &5 &6 &\bf9 & S& 8 & 5 & 26 & 6 & ? & 1\NLHL
25 & 4 &5 &7 &\bf9 &\N& 7 & 5 & 32 & 6 & ? & 1\NLHL
26 & 1 &\bf5 &7 &13 &\N& 21 & 5 & 24 & 5 & 1 & 1\NLHL
26 & 2 &3 &8 &13 &\N& 16 & 5 & 23 & 5 & 1 & 0\NLHL
26 & 2 &5 &\bf6 &13 &\N& 13 & 5 & 23 & 5 & 2 & 1\NLHL
27 & 2 &5 &\bf9 &11 &\N& 11 & 6 & 32 & 6 & 2 & 1\NLHL
27 & 5 &6 &7 &9 &\N& 6 & 5 & 30 & 6 & ? & 0\NLHL
28 & 1 &\bf4 &9 &14 &\N& 24 & 4 & 24 & 4 & 1 & 1\NLHL
28 & 3 &4 &\bf7 &14 & S& 12 & 5 & 18 & 5 & 2 & 1\NLHL
28 & 4 &6 &7 &\bf11 &\N& 7 & 4 & 35 & 4 & ? & 1\NLHL
30 & 1 &\bf4 &10 &15 & S& 25 & 5 & 20 & 5 & 1 & 1\NLHL
30 & 1 &\bf6 &8 &15 & S& 21 & 5 & 24 & 5 & 1 & 1\NLHL
30 & 2 &3 &10 &15 & S& 18 & 4 & 18 & 4 & 1 & 0\NLHL
30 & 2 &\bf6 &7 &15 &\N& 13 & 4 & 23 & 4 & 1 & 1\NLHL
30 & 3 &4 &\bf10 &13 &\N& 10 & 5 & 35 & 5 & 2 & 1\NLHL
30 & 4 &5 &6 &15 & S& 10 & 5 & 20 & 5 & ? & 0\NLHL
30 & 5 &6 &8 &\bf11 &\N& 6 & 4 & 39 & 4 & ? & 1\NLHL
32 & 2 &5 &\bf9 &16 &\N& 13 & 5 & 29 & 5 & 2 & 1\NLHL
32 & 4 &5 &7 &16 &\N& 9 & 5 & 27 & 5 & ? & 0\NLHL
33 & 3 &5 &\bf11 &14 &\N& 9 & 4 & 39 & 4 & 2 & 1\NLHL
34 & 3 &4 &\bf10 &17 &\N& 11 & 6 & 31 & 6 & 2 & 1\NLHL
34 & 4 &6 &7 &\bf17 &\N& 8 & 5 & 31 & 5 & ? & 1\NLHL
36 & 1 &\bf5 &12 &18 &\N& 24 & 4 & 24 & 4 & 1 & 1\NLHL
36 & 3 &4 &\bf11 &18 &\N& 12 & 4 & 30 & 4 & 2 & 1\NLHL
36 & 7 &8 &9 &12 &\N& 5 & 4 & 35 & 4 & ? & 0\NLHL
38 & 3 &5 &\bf11 &19 &\N& 10 & 5 & 35 & 5 & 2 & 1\NLHL
38 & 5 &6 &8 &\bf19 &\N& 7 & 5 & 35 & 5 & ? & 1\NLHL
40 & 5 &7 &8 &20 &\N& 8 & 4 & 28 & 4 & ? & 0\NLHL
42 & 1 &\bf6 &14 &21 & S& 24 & 4 & 24 & 4 & 1 & 1\NLHL
42 & 2 &5 &14 &21 &\N& 15 & 4 & 27 & 4 & 1 & 0\NLHL
42 & 3 &4 &\bf14 &21 & S& 13 & 5 & 26 & 5 & 2 & 1\NLHL
44 & 4 &5 &\bf13 &22 &\N& 9 & 4 & 39 & 4 & 2 & 1\NLHL
48 & 3 &5 &\bf16 &24 &\N& 12 & 4 & 30 & 4 & 2 & 1\NLHL
50 & 7 &8 &10 &\bf25 &\N& 6 & 4 & 39 & 4 & ? & 1\NLHL
54 & 4 &5 &\bf18 &27 &\N& 10 & 5 & 35 & 5 & 2 & 1\NLHL
66 & 5 &6 &\bf22 &33 &\N& 9 & 4 & 39 & 4 & 2 & 1\NLHL
				&&&&&&&&&&&	\NLHL
\hline\end{tabular}\hspace*{-6mm}
}
\\[4mm]
	{\bf Table 7:} 	Elliptic fibration data for the 95 K3 weights
			(58 are spanning; all are transversal)
\end{center}
\end{figure}
% Here works well.

\begin{figure}
\begin{center}				
\begin{tabular}{||c||c|c|c||c||} \hline\hline
	& \nFa=0& \nFa=1& \nFa=2& total	\\\hline\hline
\nPro=? & 7	&	11	& 0 &	18 \\\hline
\nPro=0 & 5 &  &  & 5 \\\hline
\nPro=1 & 5 & 42 &  & 47 \\\hline
\nPro=2 & 0 & 22 & 1 & 23 \\\hline
\nPro=3 & 0 & 0 & 2 & 2 \\\hline	\hline 
total & 17 & 75 & 3 & 95 \\\hline
\hline
\end{tabular}
\\[4mm]
{\bf Table 8:} 	Numbers of reflexive projections (\nPro) and facets (\nFa)
		for the K3 weights
\end{center}
\end{figure}

\begin{figure}
\begin{center}
\begin{tabular}{||c||c|c|c||c||} \hline\hline
Spectra	& 	MNPs	& no mirror 	& together	& $\c=0$\\\hline\hline
$WP^4$	&   	2,780	& 669		& 3,449		& 63	\\\hline
orbifolds & 	3,537	& 781		& 4,318		& 72	\\\hline
MNPs	&   	10,237	& 3,884		& 14,121	& 103	\\\hline\hline
total	&	10,410	& 3,806		& 14,216	& 104	\\\hline
\hline\end{tabular}
\\[4mm]
	{\bf Table 9:} Hodge numbers for transversal weights, orbifolds and
		maximal Newton polyhedra 
\end{center}
\end{figure}

\ifundefined{epsfbox}\else

\begin{figure}
\epsfxsize=15cm
\hfil\epsfbox{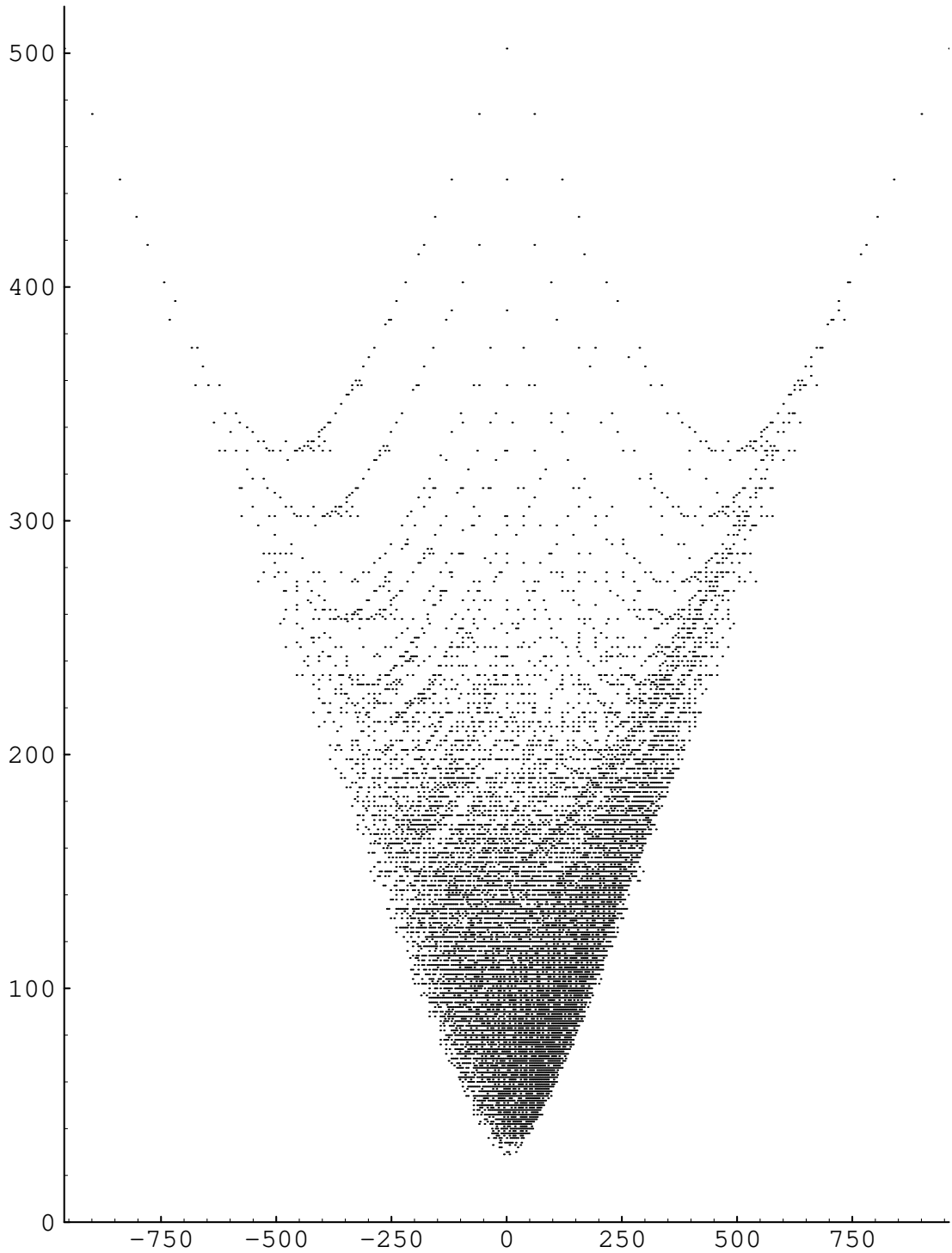}\hfill
\caption{$b_{11}+b_{21}\; vs\; \chi_E \;$ for single weight systems}
\label{fig:hdg}
\end{figure}
\newpage
\fi

\vfill\break
\bigskip	{\it Acknowledgements.}
A.C.A. is supported by the Robert A. Welch Foundation and by the NSF under 
grant PHY/9511632. M.K. is supported by the {\it Austrian Research Fund} 
under grant number P10641-PHY. M.M. is supported by the United  States 
National Science Foundation
under grants PHY-9404057 and PHY-9457916. 
H.S. is supported by the {\it Austrian National Bank} under grant number 5674. 
It is a pleasure to acknowledge useful conversations with 
Per Berglund, Philip Candelas and Anamaria Font. Most of this work was done 
while M.M. was at the University of Texas at Austin.
%\newpage

%%%%%%%%%%%%%%%%	References  ... first some junk		%%%%%%%%%%%

\del
\.ca95	P.Candelas, X. de la Ossa, S.Katz, \I{Mirror symmetry for Calabi--Yau
	hypersurfaces in weighted $\IP^4$ and extensions of Landau--Ginzburg
	theory,} hep-th/9412117, \npb 450 (1995) 267.\>
\.klun	A.Klemm, unpublished	\>
\.st95	A.Strominger, \I{Massless black holes and conifold in string theory,}
	hep-th/9504090.   \>
\.gms	B.R.Greene, D.R.Morrison, A.Strominger,
	\I{Black hole condensation and the unification of string vacua,}
	\npb 451 (1995) 109, hep-th/9504145.   \>
\.cggk	T.-M.Chiang, B.R.Greene, M.Gross, Y.Kanter,
	\I{Black hole condensation and the web of Calabi--Yau
	manifolds,} hep-th 9511204.  \>
\.acjm  A.C.Avram, P.Candelas, D.Jancic, M.Mandelberg,
	\I{On the connectedness of the moduli Space of Calabi--Yau
	manifolds,} hep-th 9511230.  \>
\.reid  M.Reid, \I{Canonical 3-folds,} Proc. Alg. Geom. Anger 1979,
        Sijthoff and Nordhoff, 273. \>
\.fl89	A.R.Fletcher, \I{Working with complete intersections,} Bonn preprint
	MPI/89--35 (1989).\>
\enddel

%%%%%%%%%%%%%%%%%%%%%%%%	References	%%%%%%%%%%%%%%%%%%%%%%%%

~\vfill
%\pagebreak[4]
%\newpage


\begin{thebibliography}{11}

\ifundefined{draftmode}	\def\.#1 #2\>{\bibitem{#1}#2}		
\else			\def\.#1 #2\>{\bibitem{#1}\LLab{#1}#2}  \fi

\def\I#1{{\it #1}}	\addtolength{\itemsep}{-4.5pt} 	\small \vspace{-3mm}

\.chsw 	P.~Candelas,~G.~Horowitz,~A.~Strominger and E.~Witten,~\I{Vacuum 
	configurations for superstrings,}\npb 258 (1985) 46. \>

\.cdls 	P.~Candelas,~A.~Dale,~C.~L\"{u}tken, and R.~Schimmrigk, \I{Complete 
	intersection Calabi-Yau manifolds,}\npb 298 (1988) 493. \>

\.yt 	S.~Yau and G.~Tian, in \I{Proc. Argonne Symposium on Anomalies, 
	Geometry and Topology,}\hfill\break ed. W.~Bardeen, A.~White, 
	World Scientific, Singapore, 1985. \>

\.nms	M.Kreuzer, H.Skarke, \I{
     	No mirror symmetry in Landau-Ginzburg spectra!,} hep-th/9205004
     	Nucl. Phys. B388 (1992) 113. 	\>

\.kl94	A.Klemm, R.Schimmrigk, \I{Landau--Ginzburg String Vacua,}
	hep-th/9204060, \npb 411 (1994) 559.	\>

\.ba94	V.V.Batyrev, \I{Dual polyhedra and mirror symmetry for Calabi--Yau
    	hypersurfaces in toric varieties,} alg-geom/9310003,
	J. Alg. Geom. {\bf 3} (1994) 493.	\>

\.crp   M.Kreuzer, H.Skarke, \I{On the classification of reflexive polyhedra,}
        hep-th/9512204, TUW-95-26.  \>

\.wtc   H.Skarke,    % TUW-96/04, alg-geom/9603007.
        \I{Weight systems for toric Calabi--Yau varieties and
         reflexivity of Newton polyhedra,}
               Mod. Phys. Lett. {\bf A11} (1996) 1637.   \>

\.kv 	S.~Kachru and C.~Vafa,~\I{Exact results for $N=2$ compactifications 
	of heterotic strings,} \npb 450 (1995) 69. \>

\.sw 	N.~Seiberg and E.~Witten,~\I{Electric-magnetic duality, monopole 
	condensation, and confinement\hfill\break in $N=2$ supersymmetric 
	Yang-Mills theory,}\npb 426 (1994) 19.\hfill\break
	N.~Seiberg and E.~Witten,~\I{Monopoles, duality, and chiral symmetry 
	breaking in $N=2$\hfill\break supersymmetric QCD,}\npb 431 (1994) 484.
	\>

\.klm 	A.~Klemm,~W.~Lerche and P.~Mayr,~\I{$K3$ fibrations and 
	heterotic-Type II String Duality,}\hfill\break \plb 357 (1995) 313. \>

\.hly   S.Hosono, B.H.Lian, S.-T. Yau,
        \I{Calabi-Yau Varieties and Pencils of K3 Surfaces,}
        \\alg-geom/9603020.  \>

\.al 	P.~Aspinwall and J.~Louis,~\I{On the ubiquity of $K3$ fibrations in 
	string duality,}\hfill\break\plb 369 (1996) 233. \>

\.cf 	P.~Candelas and A.~Font,~\I{Duality between the webs of heterotic and 
	Type II vacua,}\hfill\break hep-th/9603170, UTTG-04-96. \>

\.ful	W.Fulton, \I{Introduction to Toric Varieties,}
	Princeton University Press (1993).\>

\.cox	D.Cox, \I{The homogeneous coordinate ring or a toric variety,}
	J. Alg. Geom. {\bf 4} (1995) 17. \>

\.aud	M.Audin, \I{The Topology of Torus Actions on Symplectic Manifolds,}
	Progress in Math. {\bf 93}, Birkh\"auser, 1991.\>

\.wit93	E.Witten, \I{Phases of $N=2$ theories in two dimensions,}
	\npb 403 (1993) 159.\>

\.ao	M. Kreuzer and H. Skarke,\I{
        All Abelian Symmetries of Landau-Ginzburg Potentials,}\hfill\break 
        \npb 405 (1993) 305.    \>
%\.as96	P.S.Aspinwall, M.Gross, \I{Heterotic-heterotic string duality and
%	multiple K3 fibrations,} \plb 382 (1996) 81, hep-th/9602118. \>

\end{thebibliography}
\end{document}